\documentclass[reprint,amsmath,amssymb,aps,superscriptaddress,showpacs,floatfix,byrevtex,longbibliography]{revtex4-1}

\usepackage{amsmath}
\usepackage{amssymb}
\usepackage{amstext}
\usepackage{amsopn}
\usepackage{amsfonts}
\usepackage{amsxtra}
\usepackage[english]{babel}
\usepackage{graphicx}
\usepackage{float}
\usepackage{bm}
\usepackage{multirow}
\usepackage{dcolumn}
\usepackage{color}
\usepackage{hyperref}

\begin{document}

\preprint{Draft --- not for distribution}

%
% The title and the list of authors
%
\title{Vibrational anomalies in \emph{A}Fe$_\mathbf{2}$As$_\mathbf{2}$ (\emph{A}$\,=\,$Ca, Sr, and Ba) single crystals}
\author{C. C. Homes}
\email{homes@bnl.gov}
\affiliation{Condensed Matter Physics and Materials Science Division,
  Brookhaven National Laboratory, Upton, New York 11973, USA}%
\author{Y. M. Dai}
%\email{ymdai@nju.edu.cn}
\affiliation{Center for Superconducting Physics and Materials, Department of Physics,
 Nanjing University, 22 Hankou Road, Nanjing 210093, China}%
\author{Ana Akrap}
\affiliation{DQMP, University of Geneva, CH-1211 Geneva 4, Switzerland}
\author{S. L. Bud'ko}
\author{P. C. Canfield}
\affiliation{Ames Laboratory, U.S. DOE, and Department of Physics and Astronomy,
  Iowa State University, Ames, Iowa 50011, USA}
\date{\today}

%
% The abstract goes here
%
\begin{abstract}
The detailed behavior of the in-plane infrared-active vibrational modes has been
determined in \emph{A}Fe$_2$As$_2$ (\emph{A}$\,=\,$Ca, Sr, and Ba) above and below the
structural and magnetic transition at $T_N=172$, 195 and 138~K, respectively.
Above $T_N$, two infrared-active $E_u$ modes are observed.
In all three compounds, below $T_N$ the low-frequency $E_u$ mode is observed to split
into upper and lower branches; with the exception of the Ba material, the oscillator
strength across the transition is conserved.
In the Ca and Sr materials, the high-frequency $E_u$ mode splits into an upper and a
lower branch; however, the oscillator strengths are quite different.  Surprisingly,
in both the Sr and Ba materials, below $T_N$ the upper branch appears be either very
weak or totally absent, while the lower branch displays an anomalous increase in strength.
The frequencies and atomic characters of the lattice modes at the center of the Brillouin
zone have been calculated for the high-temperature phase for each of these materials.
The high-frequency $E_u$ mode does not change in position or character across
this series of compounds.  Below $T_N$, the $E_u$ modes are predicted to split into
features of roughly equal strength.  We discuss the possibility that the anomalous increase
in the strength of the lower branch of the high-frequency mode below $T_N$ in the Sr
and Ba compounds, and the weak (silent) upper branch, may be related to the orbital ordering
and a change in the bonding between the Fe and As atoms in the magnetically-ordered state.
\end{abstract}
%
%  PACS numbers
%  63.20.-e     Phonons in crystal lattices
%  75.30.-m 	Intrinsic properties of magnetically ordered materials
%  78.30.-j     Infrared and Raman spectra
%
\maketitle

%
% The main body of the text
%
% Introduction
%
\section{Introduction}
The discovery of superconductivity in the iron-based materials \cite{kamihara08}
with superconducting critical temperatures ($T_c$'s) in excess of 50~K \cite{ren08a,ren08b}
has prompted an intensive investigation of the physical properties of these multiband
materials in an effort to understand the pairing mechanism of the superconductivity
\cite{johnston10,paglione10,canfield10,inosov16,si16}.  The parent compounds are generally
paramagnetic metals at room temperature.  As the temperature is reduced they typically
undergo a structural and magnetic transition into antiferromagnetic (AFM) state, but
remain metallic.
The \emph{A}Fe$_2$As$_2$ (\emph{A}$\,=\,$Ca, Sr, or Ba) materials are particularly
interesting; not only do they undergo structural and magnetic transitions, but
superconductivity may also be induced through a variety of chemical substitutions, as
well as through the application of pressure.
%
% CaFe2As2
%
The compound CaFe$_2$As$_2$ is tetragonal at high temperature; depending on growth
conditions, it either undergoes a magnetic and structural transition to an orthorhombic
phase with spin-density-wave-like magnetic order at $T_N\simeq 172$~K
\cite{ni08b,goldman08,choi08,tanatar09,tanatar10}, or it undergoes a dramatic
decrease of the \emph{c}-axis lattice parameter and enters into a non-magnetic,
collapsed-tetragonal (cT) phase below 100~K \cite{canfield09a,ran11};
the cT phase, which is unique to this material, may also be stabilized by the
application of pressure \cite{kreyssig08,goldman09} or through  chemical
substitution on various sites \cite{danura11,kasahara11,saha12,jeffries12}.
Superconductivity may be induced through chemical substitution, resulting in
electron \cite{shirage08,saha09,budko13,okada14}, hole \cite{wang13}, or
isovalent \cite{kasahara11} doping; co-doping with La and P has been reported
to yield a critical temperature as high as $T_c\simeq 45$~K \cite{kudo13}.
Superconductivity may also be induced through the application of pressure
\cite{kreyssig08,torikachvili08a,torikachvili08b,lee09, canfield09a}.
%
% SrFe2As2
%
In SrFe$_2$As$_2$, the structural and magnetic transition is observed at
a somewhat higher temperature, $T_N\simeq 195$~K \cite{tegel08,zhao08,blomberg11};
superconductivity may be induced through a variety of chemical substitutions
\cite{sasmal08,chen08,goko09,saha09,shi10}, as well as pressure \cite{alireza09,kitagawa09},
with $T_c$'s in the hole-doped materials as high as $\simeq 37$~K.
%
% BaFe2As2
%
The structural and magnetic transition in BaFe$_2$As$_2$ is lower than what is
observed in the other two materials, $T_N\simeq 138$~K \cite{rotter08a};
superconductivity may be induced through the application of pressure
\cite{alireza09,ishikawa09,colombier09,yamazaki10}, or by chemical substitution
\cite{sefat08,ni08a,chu09,canfield09b,rotter08b,torikachvili08b,ni08c,
chenh09,jiang09,rullier10,thaler10,nakai10,kasahara10}, with $T_c$'s as
high as $\simeq 40$~K in the hole-doped materials.

%
% Comparison with cuprates...
%
As with the cuprates, the proximity of superconductivity to an AFM region suggests
that the superconductivity may be mediated by spin fluctuations \cite{cao08,dai08,ma09,allan14},
a notion that is supported by the argument that electron-phonon coupling in this class of
materials is too small to give rise to the high transition temperatures \cite{boeri08}.
However, the behavior of the infrared-active lattice modes can, nonetheless, be quite striking;
an example of this is the infrared phonon anomaly in BaFe$_2$As$_2$ observed below the
structural and magnetic transition \cite{akrap09,wu09,nakajima11,schafgans11,charnukha13}.
In addition, the observation of a large iron isotope effect in several iron-based
superconductors suggests an unconventional role for electron-phonon coupling
may be possible \cite{liu09}.

%
% This work
%
In this work we extend our recent investigation of the electronic properties of the
\emph{A}Fe$_2$As$_2$ (\emph{A}$\,=\,$Ca, Sr, or Ba) parent compounds \cite{dai16},
to include the detailed behavior of the in-plane infrared-active vibrational modes
above and below $T_N$.
In each of the materials, two infrared-active $E_u$ modes are observed above $T_N$.
Below the structural and magnetic transition, the low-frequency $E_u$ mode splits into
an upper and a lower branch of somewhat different strengths.  In the Ca and
Sr materials, the oscillator strength is conserved across the transition; however,
in the Ba material the strength of this mode appears to weaken slightly below $T_N$.
In the Ca and Sr materials, the high-frequency $E_u$ mode splits into an upper and
lower branch below $T_N$; the two branches have dramatically different intensities.
The oscillator strength is conserved through the transition in the Ca material,
but displays an anomalous increase in strength below $T_N$ in the Sr material.
In Ba material, the high-frequency $E_u$ mode no longer appears to split below $T_N$;
instead, it appears to undergo an abrupt decrease in frequency to become the lower
branch, with the upper branch now being either very weak or completely absent.
Furthermore, the oscillator strength of this mode increases anomalously below
$T_N$ \cite{akrap09}.  Both the high and low-frequency modes in this material
display an asymmetric line shape below the structural and magnetic transition,
indicating that the lattice modes may be coupling to spin or charge excitations.
The vibrational frequencies and atomic intensities have been calculated at the center
of the Brillouin zone for the high-temperature tetragonal phase using an \emph{ab initio}
method for the three different compounds.  The high-frequency mode involves only in-plane
Fe and As displacements that shows little change in frequency or vibrational character
across the Ca$\,\rightarrow\,$Sr$\,\rightarrow\,$Ba series.  However, the position of the
low-frequency mode decreases significantly, with the character of the low-frequency mode
shifting from an almost pure alkali-earth mode in the Ca material, to a more mixed
character in the Sr and Ba compounds, that includes a significant component from the
FeAs planes.  Surprisingly, simple empirical force-constant models predict that both
infrared-active $E_u$ modes should split into two new vibrations of roughly equal
strength, contrary to what is observed for the high-frequency $E_u$ mode.
We consider the possibility that the lack of splitting and the anomalous increase in
strength observed in the high-frequency $E_u$ mode below $T_N$ may be related to the
orbital ordering and the change in the nature of the bonding between the As and Fe
atoms in the magnetically-ordered state.

%
%%%%%%%%%%%%%%%%%%%%%%%%%%%%%%%%%%%%%%%%%%%%%%%%%%%%%%%%%%%%%%%%%%%%%%%%%%%%%%%
%
% Experiment
%
\section{Experiment}
Single crystals of $A$Fe$_2$As$_2$ (\emph{A}$\,=\,$Ca, Sr, or Ba) were grown using conventional
high-temperature solution growth techniques either out of self flux (\emph{A}$\,=\,$Ba) \cite{ni08a},
or out of Sn flux (\emph{A}$\,=\,$ Ca, Sr) \cite{yan08,ni08b,ran11}, and characterized by
x-ray scattering, electrical resistivity and magnetic susceptibility measurements.
%
% bulk physical measurements. These crystals have not been annealed.
%
The reflectance of mm-sized, as-grown crystal faces has been measured at a
near-normal angle of incidence for light polarized in the \emph{a-b} planes
over a wide frequency range from the far infrared ($\simeq 3$~meV) to the
ultraviolet ($\simeq 5$~eV) for a wide variety of temperatures above and
below $T_N$ using an \emph{in situ} evaporation technique \cite{homes93}.
Only naturally-occurring crystal faces have been used; the samples have
not been cleaved or polished.  The complex conductivity is determined
from a Kramers-Kronig analysis of the reflectance \cite{dressel-book},
the details of which have been discussed in a previous publication \cite{dai16}.

%
% Results and discussion
%
\section{Results and Discussion}
The vibrational properties of these materials are determined from the
symmetry properties of the space group.  Above $T_N$, the three materials
studied all crystalize in the tetragonal $I4/mmm$ space group.  The
irreducible vibrational representation is
\begin{equation}
  \Gamma_{\rm HT} = A_{1g} + B_{1g} + 2E_g + 2A_{2u} + 2E_u,
  \nonumber
\end{equation}
where the $E_u$ and $A_{2u}$ vibrations are infrared active in the
{\em a-b} planes and along the {\em c} axis, respectively.  Accordingly,
the two infrared active modes observed in these materials at room
temperature are assigned as $E_u$ modes.
Below $T_N$, there is a weak structural transition to an orthorhombic $Fmmm$
space group, with the irreducible vibrational representation,
\begin{equation}
  \Gamma_{\rm LT} = A_g + B_{1g} + 2B_{2g} + 2B_{3g} + 2B_{1u} + 2B_{2u} + 2B_{3u},
  \nonumber
\end{equation}
where the $B_{1u}$ modes are active along the {\em c} axis, and the orthorhombic
distortion lifts the degeneracy of the $E_u$ mode and splits it into $B_{2u} +
B_{3u}$ (active along the {\em b} and {\em a} axes, respectively) for a
total of four infrared-active modes at low temperature.  However,
{\em ab initio} studies indicate that the splitting of the $E_u$ mode in
the related LaFeAsO compound should be relatively small \cite{yildirim08},
of the order of $1.5$~cm$^{-1}$ (0.2~meV).

%
% Oscillator fits
%
In the absence of coupling to a continuum of excitations, infrared-active modes
usually display a symmetric Lorentzian profile.  However, it is possible that
coupling to either a spin or charge background may be present, so the
infrared-active vibrations have been fit using a phenomenological complex dielectric
function, $\tilde\epsilon=\epsilon_1+i\epsilon_2$, for a Fano-shaped Lorentz
oscillator \cite{damascelli96,reffit},
\begin{equation}
  \tilde\epsilon(\omega) = \frac{\Omega_0^2}{\omega_0^2-\omega^2-i\gamma_0\omega}
  \left( 1+i\frac{\omega_q}{\omega} \right)^2 +
  \left( \frac{\Omega_0\omega_q}{\omega_0\omega} \right)^2,
  \label{eq:fano}
\end{equation}
where $\omega_0$, $\gamma_0$ and $\Omega_0$ are the position, width, and strength of
the vibration, respectively, and the asymmetry is described by the dimensionless parameter
$1/q=\omega_q/\omega_0$.  The complex conductivity is $\tilde\sigma(\omega) =
\sigma_1 +i\sigma_2 = -2\pi i \omega [\tilde\epsilon(\omega) - \epsilon_\infty ]/Z_0$
(in units of $\Omega^{-1}$cm$^{-1}$); $Z_0\simeq 377$~$\Omega$ is the impedance of free
space. The complex conductivity satisfies $\tilde\sigma^\ast(\omega) = \tilde\sigma(-\omega)$.
The real and imaginary parts of the optical conductivity are then:
\begin{widetext}
%
% sigma1
%
\begin{equation}
  \sigma_1(\omega) = \frac{2\pi}{Z_0}
  \frac{\Omega_0^2\left[ \gamma_0\omega^2 - 2(\omega^2\omega_0-\omega_0^3)/q - \gamma_0\omega_0^2/q^2\right]}
   {(\omega^2 - \omega_0^2)^2+\gamma_0^2 \omega^2},
   \label{eq:s1}
%
%  \frac{\Omega_0^2\left[ q^2\gamma_0\omega^2 - 2q(\omega^2\omega_0-\omega_0^3) - \gamma_0\omega_0^2\right]}
%   {q^2 \left[\gamma_0^2 \omega^2 + (\omega^2 - \omega_0^2)^2\right]}.
\end{equation}
and
%
% sigma2
%
\begin{equation}
  \sigma_2(\omega) = \frac{2\pi}{Z_0}
  \frac{\omega\Omega_0^2\left[ (\omega^2-\omega_0^2)-2\gamma_0\omega_0/q +(\omega^2-\omega_0^2+\gamma_0^2)/q^2\right]}
   {(\omega^2 - \omega_0^2)^2+\gamma_0^2 \omega^2}.
  \label{eq:s2}
\end{equation}
\end{widetext}
Note that for finite $\omega_0$, in the $\omega_q\rightarrow 0$ (or $1/q^2\rightarrow 0)$ limit, the
dielectric function for a simple Lorentz oscillator is recovered; however, as $1/q^2$ increases the line
shape becomes increasingly asymmetric.

%
% Figure 1: low and high frequency Eu modes in CaFe2As2
%
\begin{figure*}[t]
\includegraphics[width=3.20in]{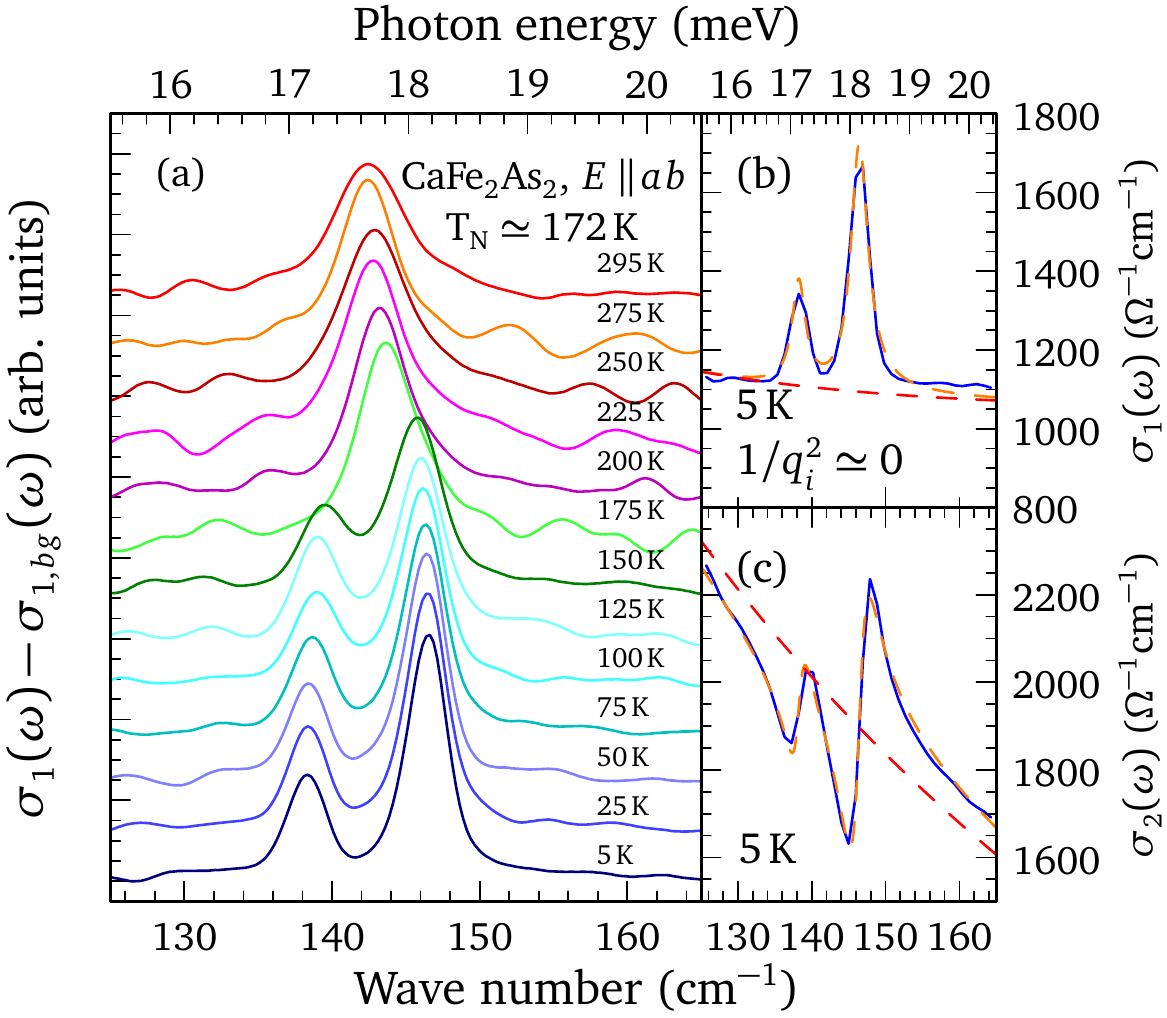}  \hspace*{6mm}
\includegraphics[width=3.20in]{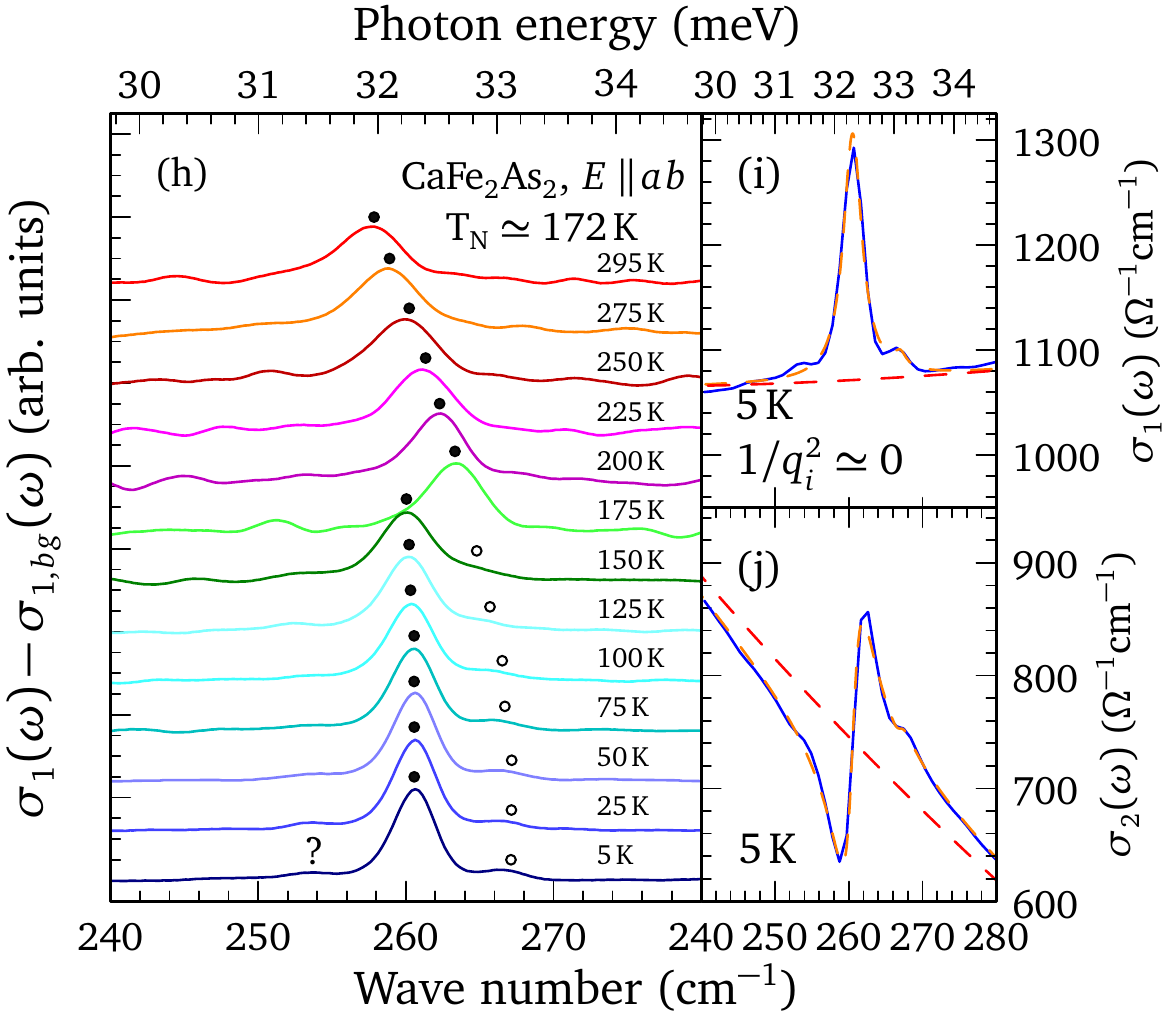} \hfil
\vspace*{2mm} \\
\hspace*{-7.0mm}
\includegraphics[width=3.34in]{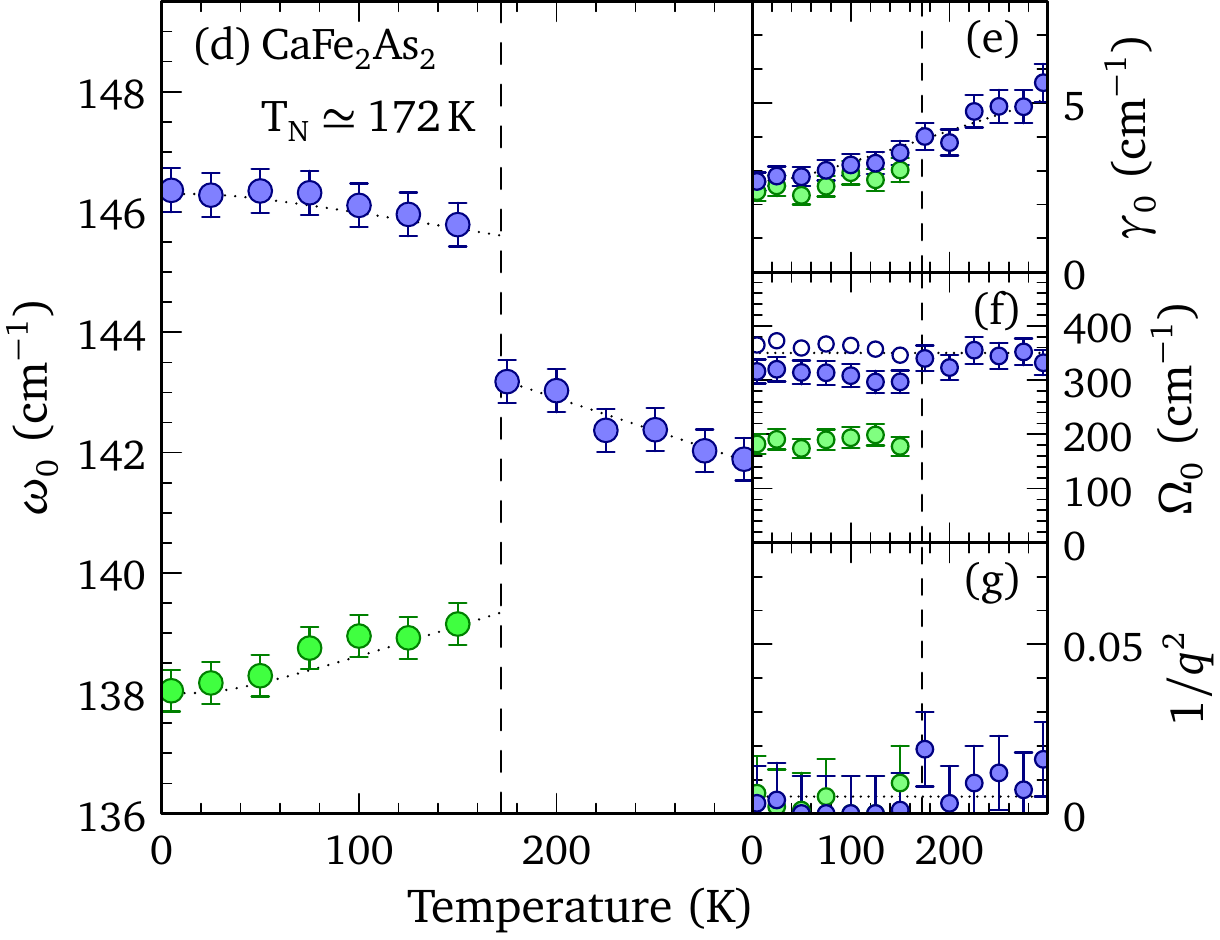} \hspace*{3mm}
\includegraphics[width=3.34in]{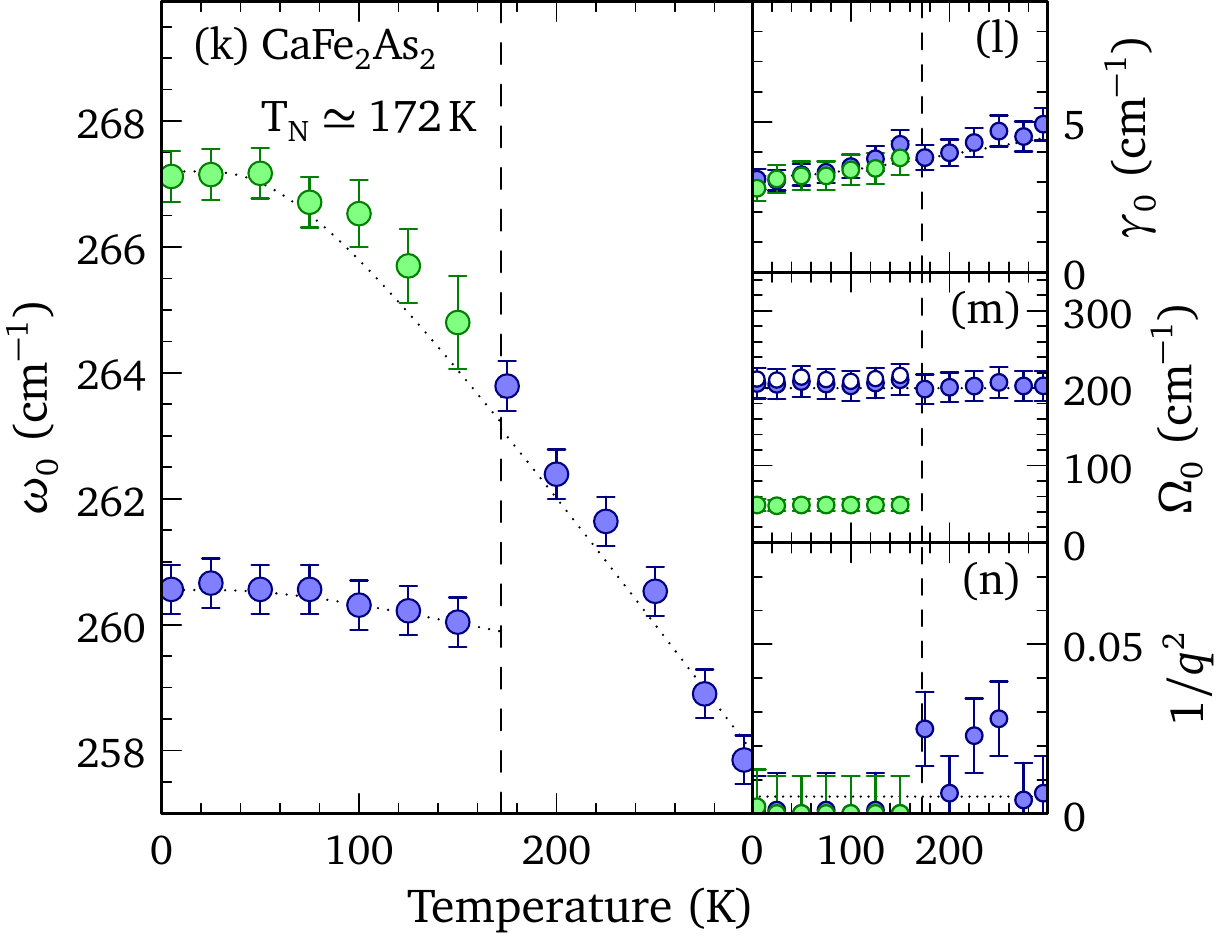} \hfil
\caption{(a) The temperature dependence of the low-frequency $E_u$ mode in the
real part of the optical conductivity for CaFe$_2$As$_2$ ($T_N\simeq 172$~K) with
the electronic background removed.  (b) The results of the fit at 5~K to the real part
of the optical conductivity, and (c) the imaginary part.
The temperature dependence of the (d) positions; (e) line widths; (f) strength of the modes
(for $T<T_N$, the open circles are the strengths of the two modes added in quadrature);
(g) asymmetry parameters.
(h) The temperature dependence of the high-frequency $E_u$ mode in the real part of
the optical conductivity with the electronic background removed; the position of the
fundamental is indicated by solid circles, below $T_N$ the position of the upper band
is denoted by the open circles.  A possible low-temperature feature of uncertain origin
is denoted by the question mark.   (i) The results of the fit at 5~K to the real part
of the optical conductivity, and (j) the imaginary part.
The temperature dependence of the (k) positions; (l) line widths; (m) strength of the modes
(for $T<T_N$, the open circles are the strengths of the two modes added in quadrature);
(n) asymmetry parameters.  Unless otherwise indicated in the text, the dotted lines are
drawn as a guide to the eye.
}
%
%\vspace*{-0.0cm}%
\label{fig:ca122}
%\end{figure}
\end{figure*}

%
% Electronic background
%
The infrared-active modes are superimposed on an electronic background, which in our
previous study of these materials \cite{dai16} was modeled using the two-Drude
model \cite{wu10}
\begin{equation}
  \tilde\epsilon(\omega) = \epsilon_\infty - \sum_{j=1}^2 {{\omega_{p,D;j}^2}\over{\omega^2+i\omega/\tau_{D,j}}}
    + \sum_k {{\Omega_k^2}\over{\omega_k^2 - \omega^2 - i\omega\gamma_k}},
  \label{eq:eps}
\end{equation}
where $\epsilon_\infty$ is the real part at high frequency.  In the
first sum, $\omega_{p,D;j}^2 = 4\pi n_je^2/m^\ast_j$ and $1/\tau_{D,j}$
are the square of the plasma frequency and scattering rate for the delocalized (Drude)
carriers in the $j$th band, respectively, and $n_j$ and $m^\ast_j$ are the carrier
concentration and effective mass.  In the second summation, $\omega_k$, $\gamma_k$ and
$\Omega_k$ are the position, width, and strength of the $k$th vibration or bound excitation.

%
%%%%%%%%%%%%%%%%%%%%%%%%%%%%%%%%%%%%%%%%%%%%%%%%%%%%%%%%%%%%%%%%%%%%%%%%%%%%%%%
%
% Table I
%
\begin{table*}[t]
\caption{Calculated frequencies and atomic intensities of \emph{A}Fe$_2$As$_2$ (\emph{A}$\,=\,$Ca, Sr, and Ba)
for the zone-center modes in the high-temperature tetragonal ($I4/mmm$) phase.}
\begin{ruledtabular}
\begin{tabular}{c ccccc c ccccc c ccccc}
   & \multicolumn{5}{c}{CaFe$_2$As$_2$}
 & & \multicolumn{5}{c}{SrFe$_2$As$_2$}
 & & \multicolumn{5}{c}{BaFe$_2$As$_2$} \\
%
% & & & \multicolumn{3}{c}{Atomic intensity} &
% & & & \multicolumn{3}{c}{Atomic intensity} &
% & & & \multicolumn{3}{c}{Atomic intensity} \\
          & $\omega_{\rm exp}$ & $\omega_{calc}$ & \multicolumn{3}{c}{Atomic intensity} &
          & $\omega_{\rm exp}$ & $\omega_{calc}$ & \multicolumn{3}{c}{Atomic intensity} &
          & $\omega_{\rm exp}$ & $\omega_{calc}$ & \multicolumn{3}{c}{Atomic intensity} \\
 Symmetry & \small{(cm$^{-1}$)} & \small{(cm$^{-1}$)} & Ca & Fe & As &
          & \small{(cm$^{-1}$)} & \small{(cm$^{-1}$)} & Sr & Fe & As &
          & \small{(cm$^{-1}$)} & \small{(cm$^{-1}$)} & Ba & Fe & As \\
\cline{1-1} \cline{2-6} \cline{8-12} \cline{13-18}
%
%                 CaFe2As2                          SrFe2As2                          BaFe2As2
% mode      wexp      wclc   Ca     Fe     As      wexp      wclc    Sr     Fe     As     wexp wclc    Ba     Fe     As
 $E_g$    & $-$     & 311 & 0.00 & 0.72 & 0.28 & & 264$^c$ & 313 & 0.00 & 0.72 & 0.28 & & 264$^a$ & 313 & 0.00 & 0.72 & 0.28 \\
 $E_u$    & 264$^a$ & 295 & 0.00 & 0.56 & 0.44 & & 257$^a$ & 294 & 0.00 & 0.55 & 0.45 & & 258$^a$ & 298 & 0.00 & 0.55 & 0.45 \\
 $A_{2u}$ & $-$     & 279 & 0.00 & 0.58 & 0.42 & & $-$     & 282 & 0.00 & 0.58 & 0.42 & & $-$     & 289 & 0.00 & 0.58 & 0.42 \\
 $B_{1g}$ & 211$^b$ & 215 & 0.00 & 1.00 & 0.00 & & 204$^c$ & 226 & 0.00 & 1.00 & 0.00 & & 217$^d$ & 237 & 0.00 & 1.00 & 0.00 \\
 $A_{1g}$ & 189$^b$ & 206 & 0.00 & 0.00 & 1.00 & & 182$^c$ & 216 & 0.00 & 0.00 & 1.00 & & 183$^d$ & 207 & 0.00 & 0.00 & 1.00 \\
 $E_g$    & $-$     & 133 & 0.00 & 0.28 & 0.72 & & 114$^c$ & 137 & 0.00 & 0.28 & 0.72 & & 131$^d$ & 146 & 0.00 & 0.28 & 0.72 \\
 $E_u$    & 143$^a$ & 127 & 0.87 & 0.07 & 0.06 & & 109$^a$ &  98 & 0.75 & 0.12 & 0.12 & &  94$^a$ &  87 & 0.65 & 0.17 & 0.18 \\
 $A_{2u}$ & $-$     & 119 & 0.86 & 0.05 & 0.09 & & $-$     &  99 & 0.75 & 0.10 & 0.15 & & $-$     &  94 & 0.66 & 0.14 & 0.20 \\
\end{tabular}
\end{ruledtabular}
\footnotetext[1] {This work; positions determined just above $T_N$.}
\footnotetext[2] {Ref.~\onlinecite{choi08}; the $A_{1g}\, (A_g)$ and $B_{1g}$ modes are only observed below $T_N$.}
\footnotetext[3] {Ref.~\onlinecite{litvinchuk08}.}
\footnotetext[4] {Refs.~\onlinecite{chauviere09,chauviere11}; the $A_{1g}\,(A_g)$ mode is only observed below $T_N$.}
\label{tab:phonons}
\end{table*}

%
%%%%%%%%%%%%%%%%%%%%%%%%%%%%%%%%%%%%%%%%%%%%%%%%%%%%%%%%%%%%%%%%%%%%%%%%%%%%%%%
%
% Ca122
%
\subsection{CaFe$_\mathbf{2}$As$_\mathbf{2}$}
%
% Low-frequency Eu mode
%
The detailed optical properties of CaFe$_2$As$_2$ have been previously determined by
us over a wide frequency range above and below $T_N$ \cite{dai16}.  The real part of
the optical conductivity is shown with the electronic background removed in Fig.~\ref{fig:ca122}(a)
in the region of the low-frequency $E_u$ mode at $\simeq 142$~cm$^{-1}$ (the curves have
been offset for clarity).  At room temperature, only one mode may be observed; below $T_N$,
this mode clearly splits into two features at $\simeq 138$ and 146~cm$^{-1}$ in response to
the orthorhombic distortion, in agreement with a previous study \cite{charnukha13}.  The fit
to the real and imaginary parts of the optical conductivity at 5~K is shown in Figs.~\ref{fig:ca122}(b)
and \ref{fig:ca122}(c), respectively, using the Fano line shapes in Eqs.~(\ref{eq:s1}) and
(\ref{eq:s2}) with an electronic background described by the two-Drude model; the line
shapes are reproduced quite well.
The position of the low-frequency $E_u$ mode is gradually increasing with decreasing temperature
until it splits below $T_N$; the upper branch increases slightly in frequency (hardens), while
the lower branch decreases (softens) as the temperature is lowered, shown in Fig.~\ref{fig:ca122}(d).
We adopt the scheme of applying the color of the fundamental to the strongest feature; just below
$T_N$ the fundamental increases abruptly to become the upper branch.
%
% The splitting at low temperature is fairly large, $\simeq 8$~cm$^{-1}$.
%
The line width shown in Fig.~\ref{fig:ca122}(e) is decreasing with temperature;
below $T_N$ there is no discontinuity as the widths of the new modes are roughly
equal and continue to decrease with temperature.  Below $T_N$ the oscillator strength
of the lower branch is less than that of the upper branch [reflected in the color scheme of
Fig.~\ref{fig:ca122}(d)].  Individually, the strengths of these modes are less than that
of the high-temperature mode, but if the strengths are added in quadrature, that is
$\Omega_1^2+\Omega_2^2$, then they reproduce the strength of the original $E_u$ mode,
as shown in Fig.~\ref{fig:ca122}(f).  The small value of the asymmetry parameter in
Fig.~\ref{fig:ca122}(g) indicates that the modes are all symmetric, both above
and below $T_N$.

%
% High-frequency Eu mode
%
The real part of the optical conductivity is shown with the electronic background removed in
Fig.~\ref{fig:ca122}(h) in the region of the high-frequency $E_u$ mode at $\simeq 258$~cm$^{-1}$
(the curves have been offset for clarity).  At room temperature a single mode is observed;
below $T_N$ two new modes appear at $\simeq 261$ and 267~cm$^{-1}$; the fit to the real and
imaginary parts of the optical conductivity at 5~K using two oscillators reproduces the data
quite well, as shown in Figs.~\ref{fig:ca122}(i) and \ref{fig:ca122}(j), respectively.
We note that there is some evidence for an extremely weak feature at $\simeq 254$~cm$^{-1}$
below about 50~K; however, because the $E_u$ mode is doubly degenerate and is therefore
expected to yield no more than two new features, and because there are no nearby Raman modes
(Table~\ref{tab:phonons}) that might possibly be activated through a symmetry-breaking
process, the origin of this structure remains uncertain.  In general, where there
are a number of other weak features in Figs.~\ref{fig:ca122}(a) and \ref{fig:ca122}(h) that
might reasonably be argued to be due to vibrational structure, we limit our attention to
only those features that show a systematic temperature dependence.
%
% At low temperature, the separation of $\sim 6$~cm$^{-1}$ between the two branches is
% similar to that observed for the low-frequency mode.
%
Above $T_N$ the frequency of the $E_u$ mode increases relatively quickly with decreasing temperature;
below $T_N$ this mode continues as a very weak upper branch that continues to harden, while
the fundamental undergoes an abrupt decrease in frequency to form a strong lower branch that
hardens slightly with decreasing temperature, shown in Fig.~\ref{fig:ca122}(k).  Interestingly,
the dotted line connecting the fundamental above $T_N$ to the weak upper branch below $T_N$
is the expected behavior for a symmetric anharmonic decay into two acoustic modes with
identical frequencies and opposite momenta \cite{klemens66,menendez84,homes16}.
The width of the mode shown in Fig.~\ref{fig:ca122}(l) is decreasing with temperature;
below $T_N$ the width of the upper branch and lower branches are similar, and both continue
to decrease with temperature; the dotted line again indicates the behavior expected from an
anharmonic decay scheme.  The strength of the upper branch is much less than that of the lower
branch, as shown in Fig.~\ref{fig:ca122}(m), but once again when the oscillator strengths are
added in quadrature then they appear to reproduce the strength of the $E_u$ mode at high temperature.
The asymmetry parameter in Fig.~\ref{fig:ca122}(n) indicates that the modes are symmetric,
both above and below $T_N$.

%
% Figure 2: SrFe2As2
%
\begin{figure*}[t]
\includegraphics[width=3.20in]{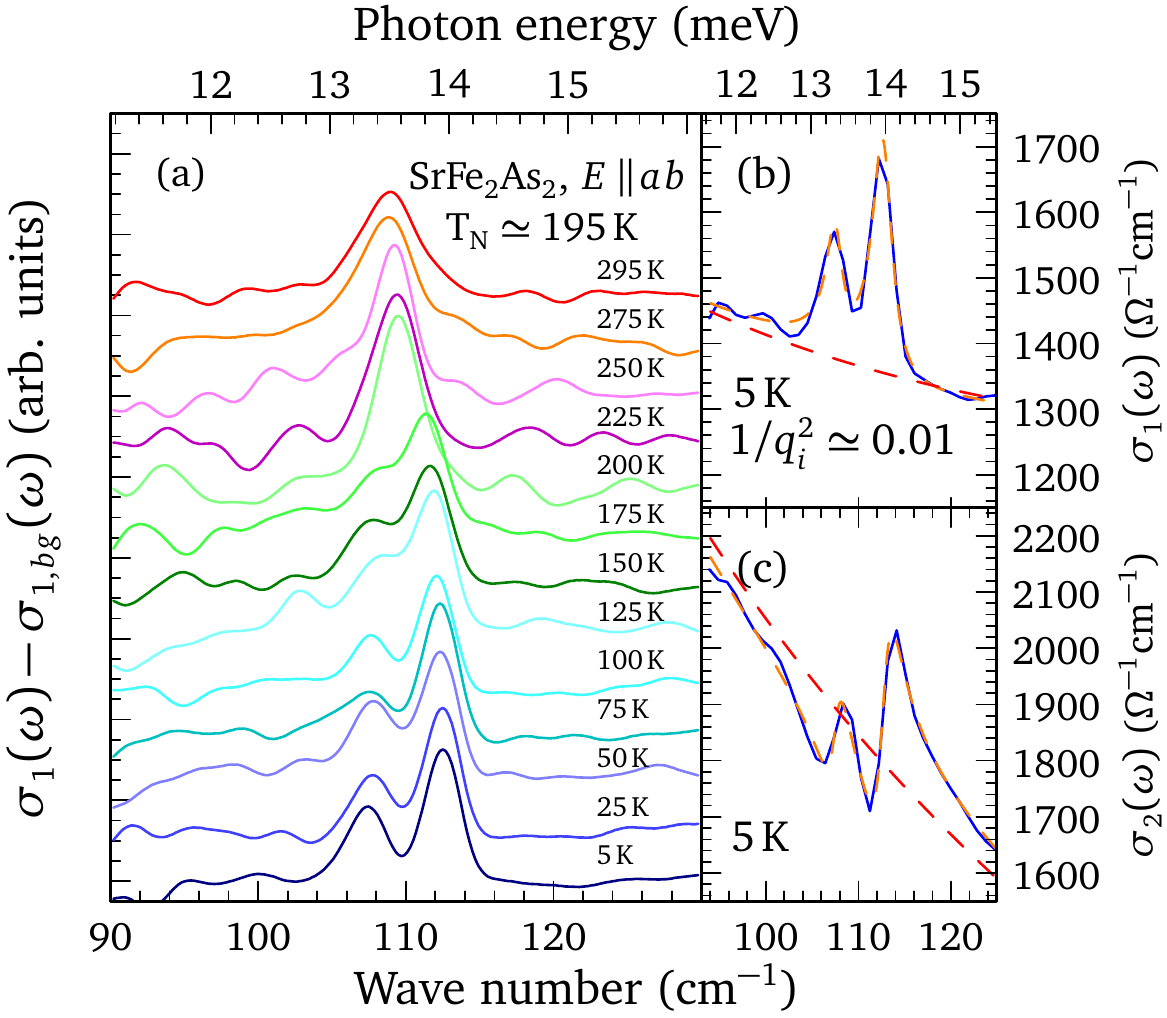} \hspace*{6mm}
\includegraphics[width=3.20in]{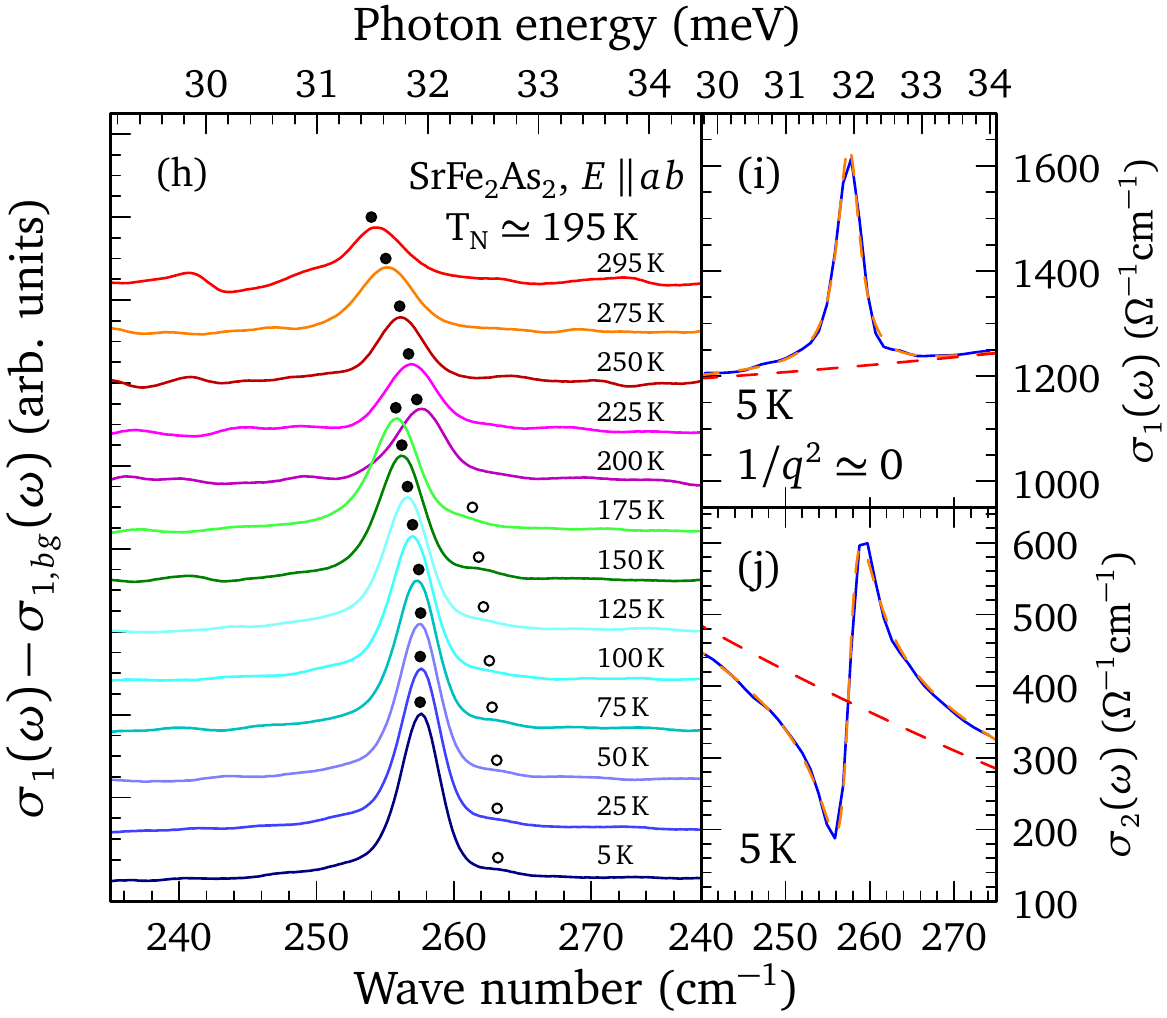} \hfil
\vspace*{2mm} \\
\hspace{-7.0mm}
\includegraphics[width=3.34in]{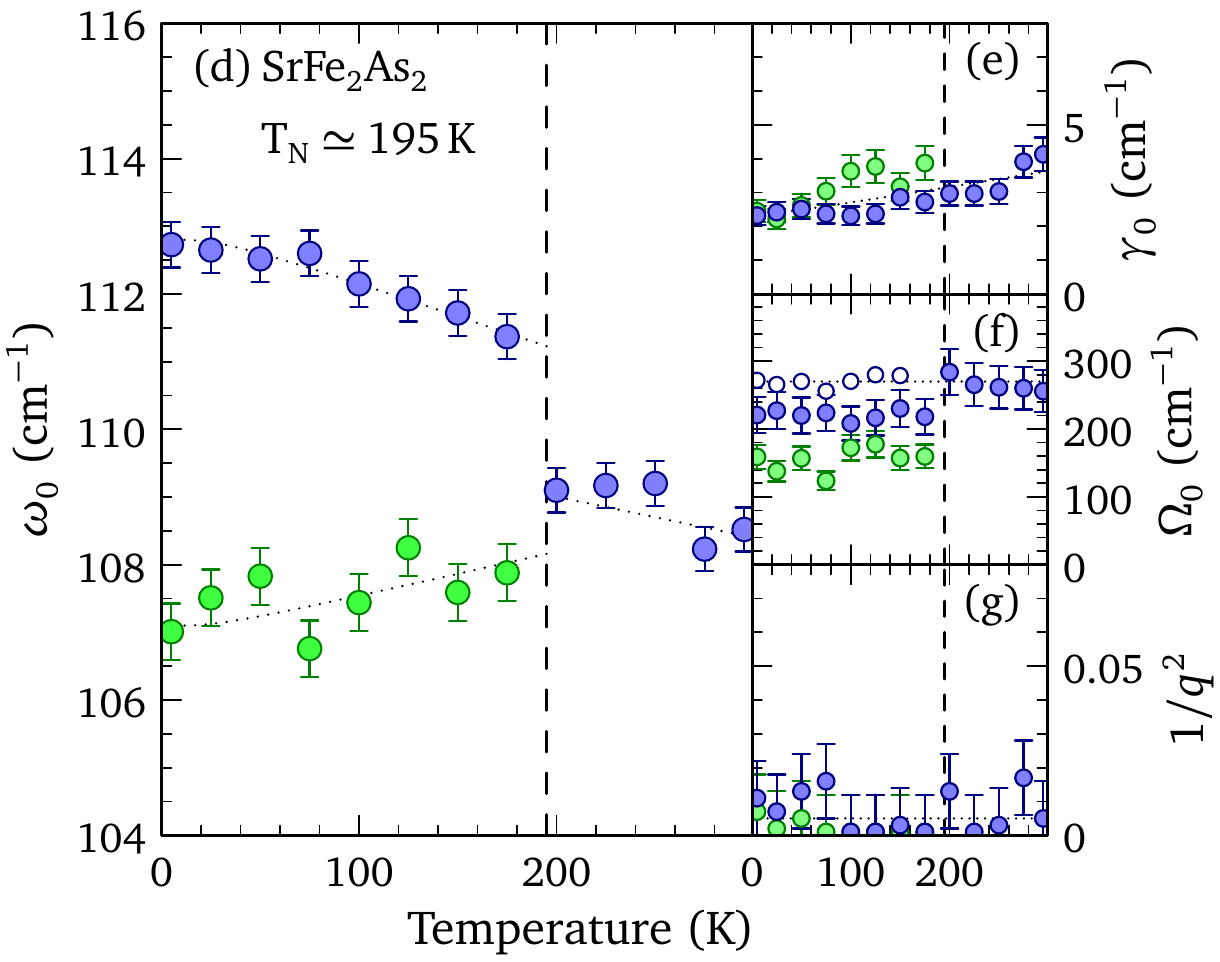} \hspace*{3mm}
\includegraphics[width=3.34in]{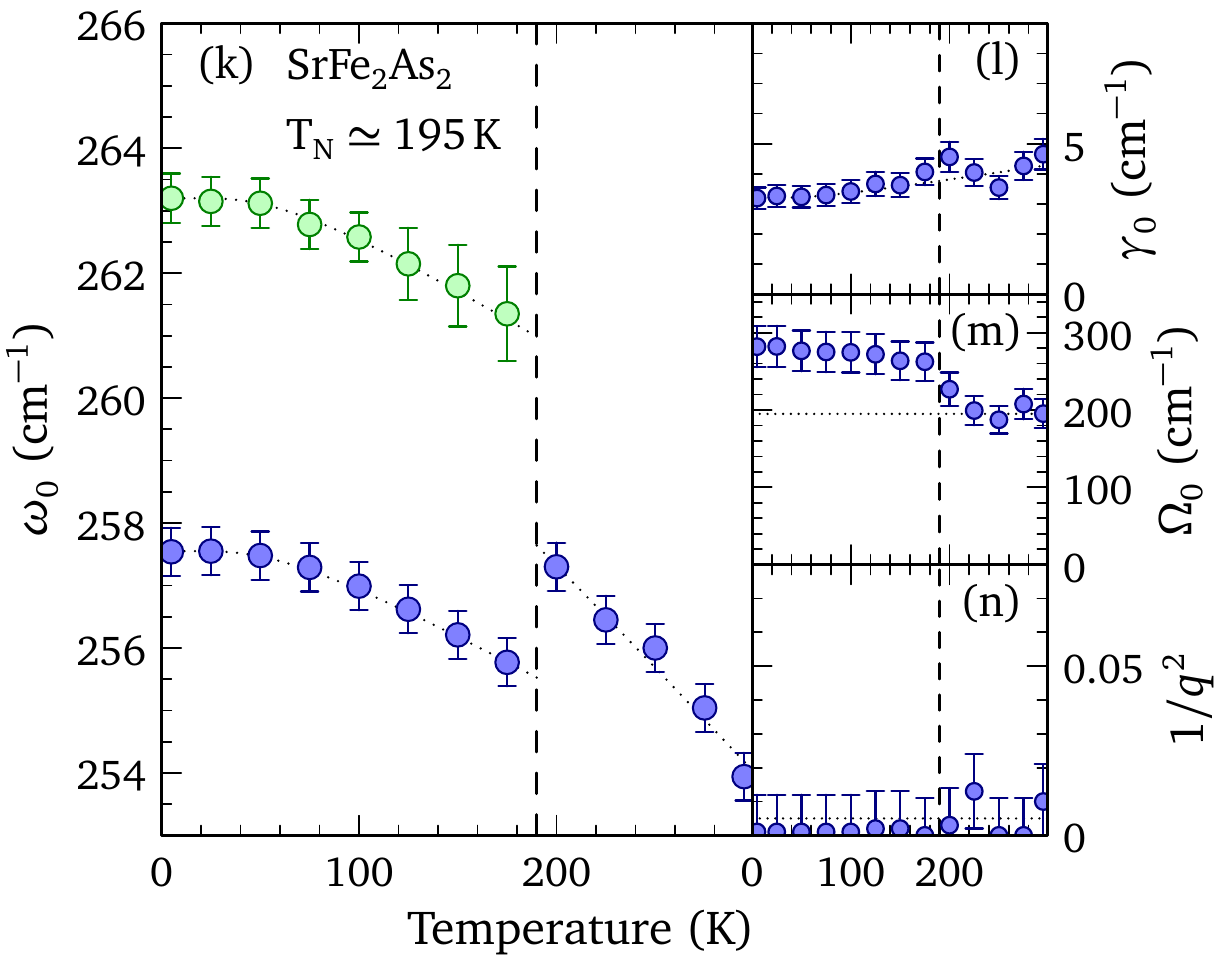} \hfil
\caption{(a) The temperature dependence of the low-frequency $E_u$ mode in the
real part of the optical conductivity for SrFe$_2$As$_2$ ($T_N\simeq 195$~K) with
the electronic background removed.  (b) The results of the fit at 5~K to the real part
of the optical conductivity, and (c) the imaginary part.
The temperature dependence of the (d) positions; (e) line widths; (f) strength of the modes
(for $T<T_N$, the open circles are the strengths of the two modes added in quadrature);
(g) asymmetry parameters.
(h) The temperature dependence of the high-frequency $E_u$ mode in the real part of
the optical conductivity with the electronic background removed; the position of the
fundamental is indicated by solid circles, below $T_N$ the position of the upper band
is denoted by the open circles.  (i) The results of the fit at 5~K to the real part
of the optical conductivity, and (j) the imaginary part.
The temperature dependence of the (k) position; (l) line width; (m) oscillator strength;
(n) asymmetry  parameter.  Unless otherwise indicated in the text, the dotted lines are
drawn as a guide to the eye.
}
%
%\vspace*{-0.0cm}%
\label{fig:sr122}
%\end{figure}
\end{figure*}

%
% Ca122 lattice dynamics
%
The frequencies and atomic intensities (the square of the vibrational amplitude of each
atom for a normal mode) have been calculated using the relaxed unit cell parameters for
the high-temperature $I4/mmm$ tetragonal phase (for details refer to the Appendix and
Table~\ref{tab:gga}); the results are shown in Table~\ref{tab:phonons}.  The position
of the low-frequency $E_u$ mode, $\simeq 143$~cm$^{-1}$ measured just above $T_N$, is
in reasonable agreement with the calculated value of 127~cm$^{-1}$; the atomic intensities
indicate that this mode involves mainly the Ca atom, with only minor contributions from
the Fe and As atoms.  The calculated position of the high-frequency mode at 295~cm$^{-1}$ is
somewhat above the observed value at 264~cm$^{-1}$; the atomic intensities indicate that
this in-plane mode involves roughly equal contributions from the Fe and As atoms, with
no contribution from the Ca atom.

%
%%%%%%%%%%%%%%%%%%%%%%%%%%%%%%%%%%%%%%%%%%%%%%%%%%%%%%%%%%%%%%%%%%%%%%%%%%%%%%%
%
\subsection{SrFe$_\mathbf{2}$As$_\mathbf{2}$}
%
% Low-frequency E_u mode
%
The detailed optical properties of SrFe$_2$As$_2$ have been previously determined by
us over a wide frequency range above and below $T_N$ \cite{dai16}.  The real part of
the optical conductivity is shown in Fig.~\ref{fig:sr122}(a), with the electronic background
removed, in the region of the low-frequency $E_u$ mode at $\simeq 109$~cm$^{-1}$ (the curves
have been offset for clarity).  At room temperature, only a single mode is observed; this
crystal is somewhat smaller than the other two, consequently the data is a bit noisier
in the long-wavelength region.  Below $T_N \simeq 195$~K this mode clearly splits into
two features at $\simeq 107$ and 112~cm$^{-1}$.  Using the previously described approach,
the Fano line shape is fit to the real and imaginary parts of the optical conductivity at
5~K as shown in Figs.~\ref{fig:sr122}(b) and \ref{fig:sr122}(c), respectively; the line
shapes are reproduced quite well.
As a result of the rather large value for $T_N$ in this material, the position of the
low-frequency $E_u$ mode shows relatively little temperature dependence before splitting
abruptly below the structural and magnetic transition; the upper branch continues to
harden with decreasing temperature, while the lower branch softens, shown in
Fig.~\ref{fig:sr122}(d).
%
% The splitting at low temperature is $\simeq 6$~cm$^{-1}$, smaller than in CaFe$_2$As$_2$.
%
The line width shown in Fig.~\ref{fig:sr122}(e) is decreasing with temperature; below $T_N$
there is no discontinuity as the widths of the new modes are roughly equal and continue to
decrease with temperature.  Below $T_N$ the oscillator strength of the lower branch is
less than that of the upper branch, but if the strengths of these two modes are added
in quadrature, they reproduce the strength of the fundamental mode, as shown in
Fig.~\ref{fig:sr122}(f).  The asymmetry parameter in Fig.~\ref{fig:sr122}(g) indicates
that the modes are all symmetric, both above and below $T_N$.

%
% Figure 3: low and high frequency Eu mode in the Ba122 materials
%
\begin{figure*}[t]
%\hspace*{2.0mm}
\includegraphics[width=3.20in]{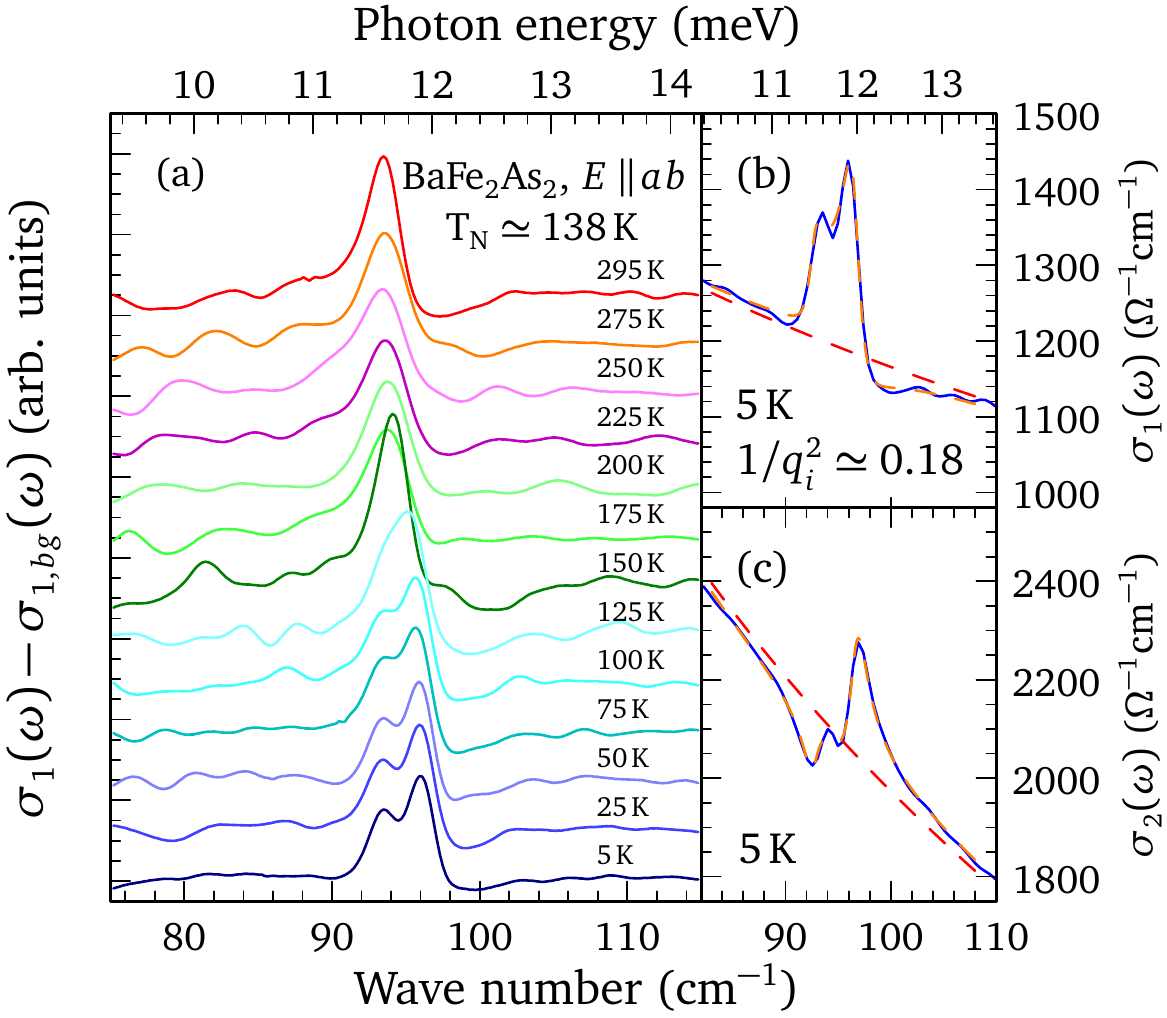} \hspace*{6mm}
\includegraphics[width=3.20in]{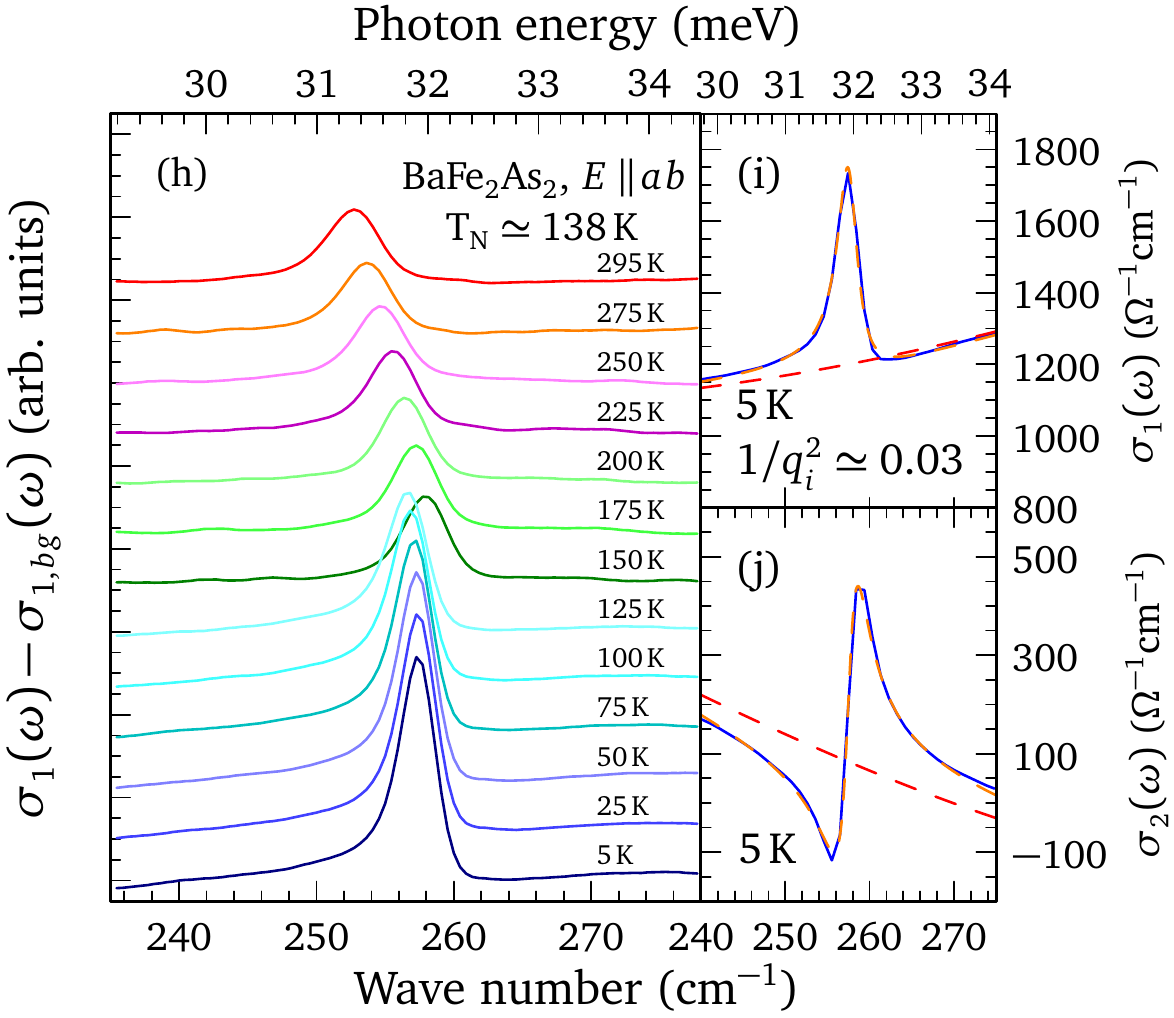} \hfil
\vspace*{2mm} \\
\hspace{-7.0mm}
\includegraphics[width=3.34in]{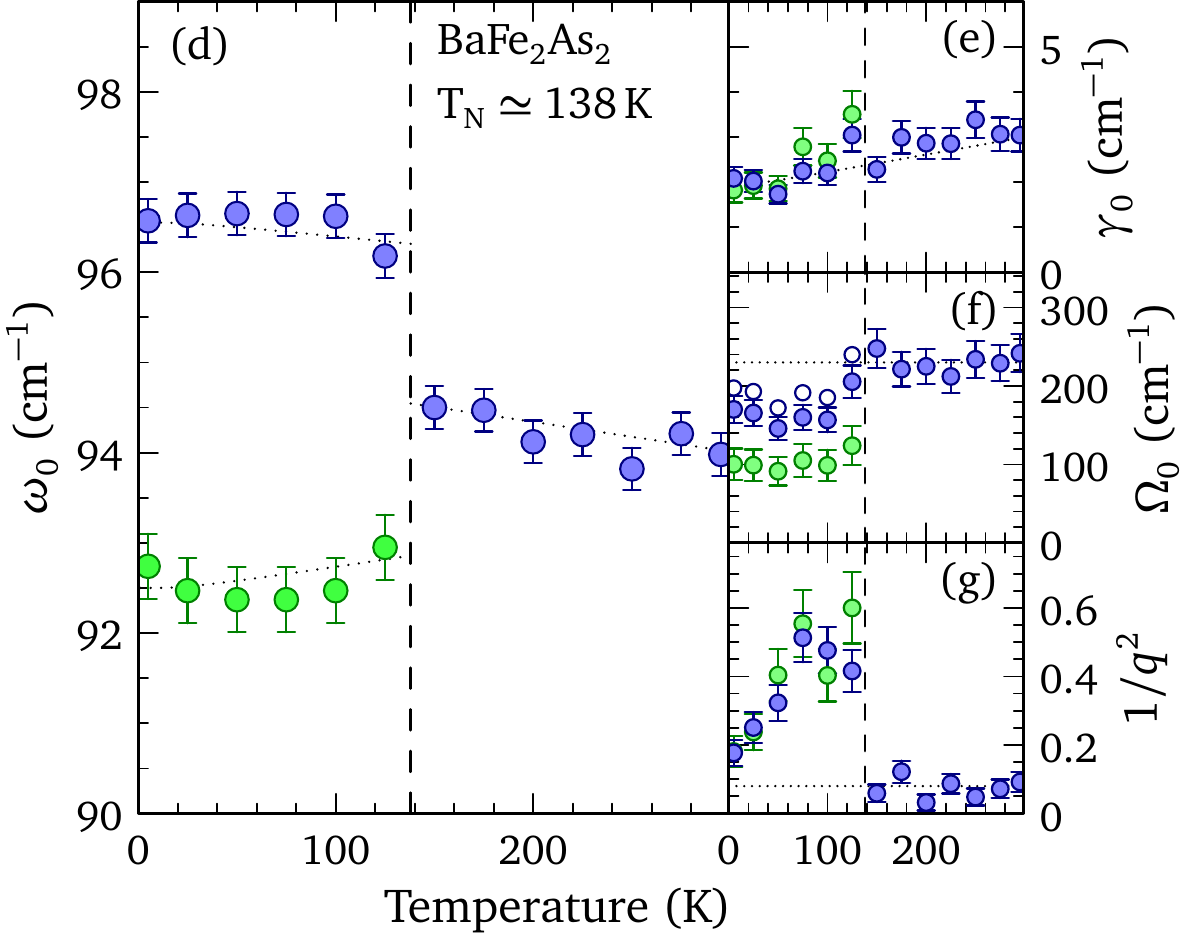} \hspace*{3mm}
\includegraphics[width=3.34in]{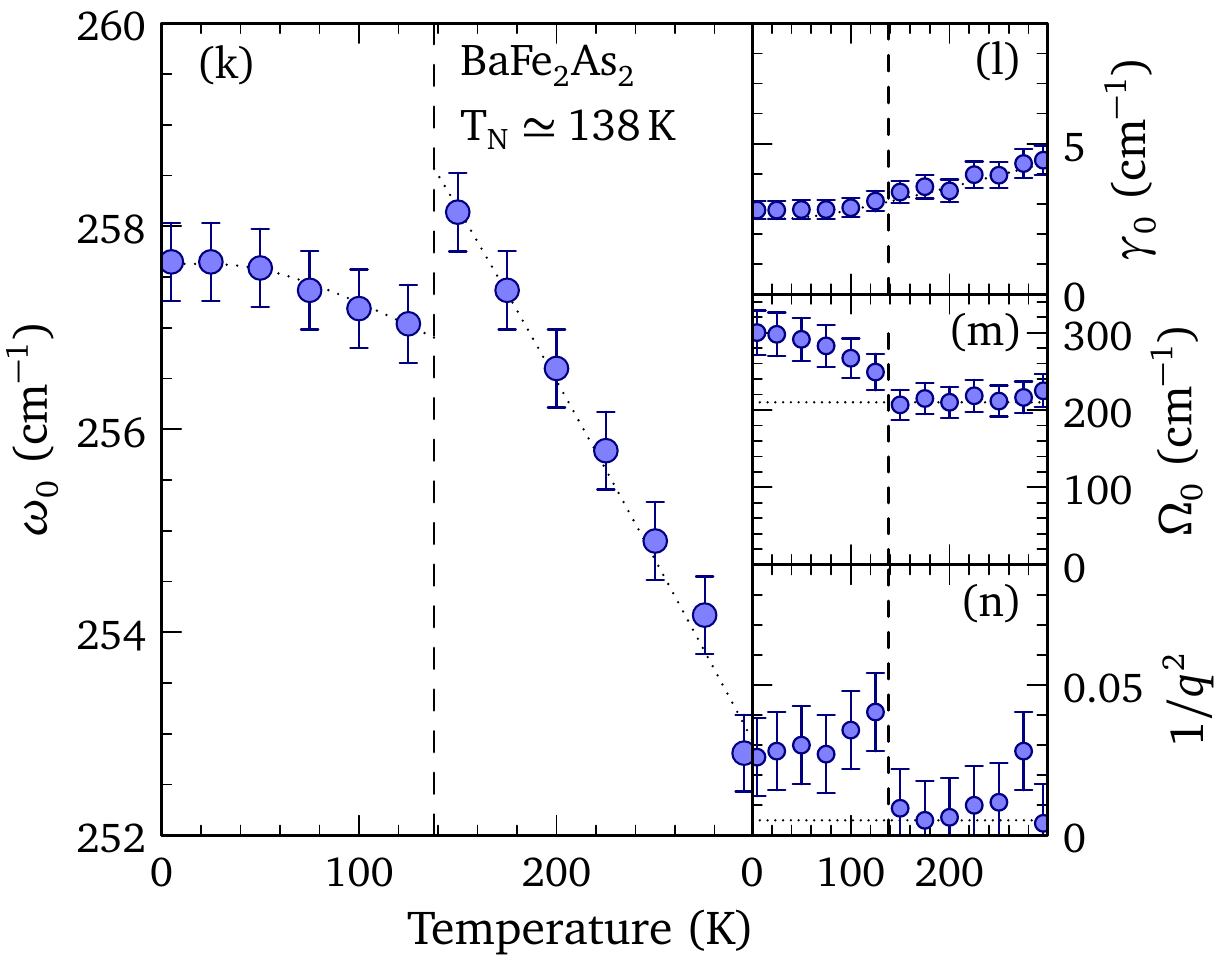} \hfil
\caption{(a) The temperature dependence of the low-frequency $E_u$ mode in the
real part of the optical conductivity for BaFe$_2$As$_2$ ($T_N\simeq 138$~K) with
the electronic background removed.  (b) The results of the fit at 5~K to the real part
of the optical conductivity, and (c) the imaginary part.
The temperature dependence of the (d) positions; (e) line widths; (f) strength of the modes
(for $T<T_N$, the open circles are the strengths of the two modes added in quadrature);
(g) asymmetry parameters.
(h) The temperature dependence of the high-frequency $E_u$ mode in the real part of
the optical conductivity with the electronic background removed.  (i) The results of
the fit at 5~K to the real part of the optical conductivity, and (j) the imaginary part.
The temperature dependence of the (k) position; (l) line width; (m) oscillator strength;
(n) asymmetry parameter.  Unless otherwise indicated in the text, the dotted lines are
drawn as a guide to the eye.
}
%
%\vspace*{-0.0cm}%
\label{fig:ba122}
%\end{figure}
\end{figure*}

%
% High-frequency Eu mode
%
The real part of the optical conductivity is shown with the electronic background removed in
Fig.~\ref{fig:sr122}(h) in the region of the high-frequency $E_u$ mode at $\simeq 254$~cm$^{-1}$
(the curves have been offset for clarity). For $T\ll T_N$, there is some evidence that this mode
splits, but the shoulder just above the main peak is extremely weak and difficult to refine. As a
consequence, we have used only a single oscillator to fit the data, while estimating the position
of the shoulder from the peak in the conductivity (the width and oscillator strength of this
feature are not determined).  The fit to the real and imaginary parts of the optical
conductivity at 5~K using the Fano line shape reproduces the data quite well, as shown in
Figs.~\ref{fig:sr122}(i) and \ref{fig:sr122}(j), respectively.
Above $T_N$ the position of this mode hardens with decreasing temperature before splitting
below $T_N$ into a strong lower branch and a weak upper branch, both of which continue to
harden with decreasing temperature, shown in Fig.~\ref{fig:sr122}(k).  The width the mode
shown in Fig.~\ref{fig:sr122}(l) is decreasing continuously with temperature and shows no
anomalous behavior at $T_N$.  The temperature dependence of the frequency and position is
consistent with the anharmonic decay scheme (dotted lines).
Above $T_N$, the strength of this mode is identical to that observed in CaFe$_2$As$_2$; however,
below $T_N$ this mode displays an unusual increase in strength.  The asymmetry parameter in
Fig.~\ref{fig:sr122}(n) indicates that the mode is symmetric, both above and below $T_N$.

%
% Lattice dynamics
%
The calculated position of the high-frequency $E_u$ mode for SrFe$_2$As$_2$ in
the tetragonal phase at 294~cm$^{-1}$ is once again somewhat above the
experimentally-observed position at $\simeq 257$~cm$^{-1}$ (taken just above
$T_N$); however, the nature of this mode is unchanged from CaFe$_2$As$_2$ with
the atomic intensities nearly identical (Table~\ref{tab:phonons}).  The
calculated position of the low-frequency $E_u$ mode at 98~cm$^{-1}$ is significantly
lower than the value of 127~cm$^{-1}$ calculated for CaFe$_2$As$_2$, and is in
reasonable agreement with the experimentally-observed position at 109~cm$^{-1}$;
the character of this mode is altered with the Sr atom now playing a decreased
role, and larger contributions from the Fe and As atoms.

%
%%%%%%%%%%%%%%%%%%%%%%%%%%%%%%%%%%%%%%%%%%%%%%%%%%%%%%%%%%%%%%%%%%%%%%%%%%%%%%%
%
% Ba122
%
\subsection{BaFe$_2$As$_2$}
%
% Low-frequency E_u mode
%
The detailed optical properties of BaFe$_2$As$_2$ have been previously determined by
us over a wide frequency range above and below $T_N$ \cite{dai16}.  The real part of
the optical conductivity is shown in Fig.~\ref{fig:ba122}(a), with the electronic
background removed, in the region of the low-frequency $E_u$ mode at $\simeq 94$~cm$^{-1}$
(the curves have been offset for clarity).  As in the other two cases, at room temperature
a single mode is observed; below $T_N$ it splits into two modes at $\simeq 93$ and
97~cm$^{-1}$.  Using the previously described approach, the Fano line shape
is fit to the real and imaginary parts of the optical conductivity at 5~K and is shown in
Figs.~\ref{fig:ba122}(b) and \ref{fig:ba122}(c), respectively, reproducing the data
quite well.
The position of the low-frequency $E_u$ mode above and below $T_N$ is shown in
Fig.~\ref{fig:ba122}(d); overall, this mode shows little frequency dependence, in
agreement with a previous study \cite{schafgans11}.
%
% The splitting of the modes below $T_N$ is $\simeq 4$~cm$^{-1}$.  If the ratio of the
% splitting ($\delta\omega_0)$ to the fundamental $\omega_0$ is considered, then
% $\delta\omega_0/\omega_0 \simeq 0.04$ in this material is less than the value of $0.06$
% observed in the other two materials.
%
The line width shown in Fig.~\ref{fig:ba122}(e) is decreasing with temperature;
below $T_N$ there is no discontinuity as the widths of the new modes are roughly equal
and continue to decrease with temperature.  Below $T_N$ the oscillator strength of
the lower branch is less than that of the upper branch.  Unlike the other two materials,
when the strengths of the two modes are added in quadrature, they fail to reproduce the
strength of the fundamental mode, as shown in Fig.~\ref{fig:ba122}(f).  The asymmetry
parameter in Fig.~\ref{fig:ba122}(g) indicates this mode is symmetric at high temperature;
however, below $T_N$ both modes display a pronounced asymmetry.

%
% High-frequency Eu mode
%
The real part of the optical conductivity is shown with the electronic background removed in
Fig.~\ref{fig:ba122}(h) in the region of the high-frequency $E_u$ mode at $\simeq 254$~cm$^{-1}$
(the curves have been offset for clarity).   At high temperature a single mode is observed;
below $T_N$, unlike the previous two materials, we fail to observe an unambiguous splitting.
Other workers have observed a weak shoulder in optical conductivity at $\simeq 262$~cm$^{-1}$
which they attribute to the splitting of this mode \cite{schafgans11}; however, we consider
only a single oscillator.  The fit to the real and imaginary parts of the optical conductivity
at 5~K using the Fano line shape reproduces the data quite well, as shown in Figs.~\ref{fig:ba122}(i)
and \ref{fig:ba122}(j), respectively.  The position of this mode hardens with decreasing
temperature until it softens anomalously at $T_N$ before continuing to harden slightly at
low temperature.  The softening is understood as a splitting in response to the orthorhombic
distortion in which the upper branch is curiously silent.  The width of the mode, shown in
Fig.~\ref{fig:ba122}(l), decreases with temperature, showing no discontinuity at $T_N$.
The temperature dependence of the frequency and position are both consistent with the
anharmonic decay scheme (dotted lines).
Above $T_N$, the strength of this mode is identical to that observed in the other two materials;
however, below $T_N$ this mode increases dramatically in strength, shown in Fig.~\ref{fig:ba122}(m).
The asymmetry parameter in Fig.~\ref{fig:ba122}(n) indicates that the mode is symmetric above $T_N$,
but develops a slight asymmetry at low temperature.

%
% Lattice dynamics
%
The calculated position of the low-frequency $E_u$ mode at 87~cm$^{-1}$ is in good agreement
with the experimentally observed position at $\simeq 94$~cm$^{-1}$; however, in a continuation
of the trend observed in SrFe$_2$As$_2$, the character of this mode has changed significantly
with the Ba atom now playing a reduced role with increased contributions from the Fe and As
atoms (Table~\ref{tab:phonons}).  The calculated position of the high-frequency mode at
298~cm$^{-1}$ is in reasonable agreement with the experimentally observed value at 258~cm$^{-1}$;
interestingly, the vibrational character for this mode remains unchanged across all three
materials.

\subsection{Phonon anomalies}
In \emph{A}Fe$_2$As$_2$, the series of \emph{A}$=\,$Ca, Sr and Ba represents the steadily
increasing size of the alkali earth atom, which is reflected by the increase in the
size of the unit cell (Table~\ref{tab:gga}).  The vibrational properties of CaFe$_2$As$_2$
are an interesting point of departure in that they may be described as more or less
conventional; both $E_u$ modes harden with decreasing temperature as expected \cite{klemens66,
menendez84,homes16} and the $E_u\rightarrow B_{2u}+B_{3u}$ splitting is observed below $T_N$
for both modes.  While one mode is typically stronger than the other, when the strengths
of the modes are added in quadrature, the strength of the original $E_u$ mode is recovered.
%
%Following the strengths of the $E_u$ modes through the transition, it is clear that in
%response to the orthorhombic distortion below $T_N$, the low-frequency mode continues as
%the upper branch, while the high-frequency mode continues as the lower branch.
%
It is not unusual for the strengths of the $B_{2u}$ and $B_{3u}$ modes to be different;
however, in the case of the high-frequency $E_u$ mode they are significantly different,
$\Omega_2^2/\Omega_1^2 \simeq 0.05$ [Fig.~\ref{fig:ca122}(m)].  This is rather
surprising as a simple empirical force-constant model \cite{vibratz} that reproduces the
positions of the vibrations below $T_N$ indicates that while there is some difference in
the strengths of the low-frequency modes, the strengths of the high-frequency $B_{2u}$ and
$B_{3u}$ modes should be nearly identical.
%
% SrFe2As2
%
In SrFe$_2$As$_2$, the low-frequency $E_u$ mode hardens with decreasing temperature and
underdoes a $B_{2u}+B_{3u}$ splitting below $T_N$; the oscillator strength is
conserved.  Beyond this point, any similarity with the Ca compound vanishes;
while the high-frequency $E_u$ mode appears to split below $T_N$, the response is
dominated by the lower branch, while the shoulder that  comprises the upper branch
is extremely weak.  In addition, the strength of the lower branch is observed to
increase anomalously below $T_N$ [Fig.~\ref{fig:sr122}(m)]  by nearly a factor of two.

%
% BaFe2As2
%
The most striking phonon anomalies occur in BaFe$_2$As$_2$.  While the
splitting of the low-frequency $E_u$ mode is observed below $T_N$ \cite{schafgans11},
the strengths of the new modes now sum to less than the original $E_u$ mode
[Fig.~\ref{fig:ba122}(f)].  Additionally, both modes appear to have asymmetric line
shapes; considering that these modes have an increased Fe and As atomic character,
the possibility of lattice mode coupling to spin or charge excitations within the
conducting FeAs planes can not be discounted.    Furthermore, the high-frequency
$E_u$ mode fails to demonstrate any splitting below $T_N$ and undergoes a dramatic
increase in oscillator strength of $\simeq 2.3$ \cite{akrap09,schafgans11}.
Below $T_N$ the spins order antiferromagnetically along the \emph{a} axis and
ferromagnetically along the \emph{b} axis \cite{goldman08,kofu09}.  Optical studies
of mechanically-detwinned single crystals of this compound indicate that there is a
significant optical anisotropy with the optical conductivity along the \emph{a} axis
being roughly twice that along the \emph{b} axis over much of the infrared region
\cite{dusza11,dusza12,nakajima11}.  The high-frequency mode is only observed
along the \emph{b} axis below $T_N$, and is thus a $B_{2u}$ mode \cite{nakajima11}.
The absence of the $B_{3u}$ mode has been attributed to screening effects
\cite{schafgans11}; however, given that the conductivity anisotropy in the region
of the mode is only about a factor of two, screening seems unlikely \cite{homes00}.

The bonding in the highly-distorted FeAs planes occurs from the hybridization of
the Fe $d_{xz}$ and $d_{yz}$ and As $4p$ orbitals \cite{li08b,andersen11}; below
$T_N$ there is a strongly orbital-dependent reconstruction of the electronic
structure \cite{shimojima10,richard10,yin11b}, resulting in orbital ordering
\cite{lv09,leecc09,kruger09,valenzuela10,dai12,liang13,fernandes14,chubukov16}.
In this scenario, degeneracy between the $d_{xz}$ and $d_{yz}$ orbitals is
removed, resulting in unequal occupations.  The strength of an infrared-active
mode is calculated from the Born effective charge on the atoms, which when taken
with the atomic displacements for a particular normal mode, is used to calculate
the strength of the dipole moment.  Changes in the strength of a mode require
changes in either bonding or coordination; the weak nature of the structural
phase transition rules out the latter, leaving only changes in bonding as a
reasonable explanation.  Because there is a spin-density wave (SDW) along the
\emph{b} axis, this also implies that there is a weak bond-centered charge-density
wave (CDW) in this direction as well.  It is possible that the different charge
densities associated with the bonds may result in dramatically different dipole
moments along the \emph{a} and \emph{b} directions in the magnetically-ordered
state, resulting in the increase in strength of the $B_{2u}$ mode and the almost
total extinction of the $B_{3u}$ mode.

%
% Table II
%
\begin{table*}[t]
\caption{The experimental and theoretical lattice constants and atomic fractional
coordinates for the relaxed structure of \emph{A}Fe$_2$As$_2$ (\emph{A}$\,=\,$Ca, Sr, and Ba)
for the high-temperature tetragonal $I4/mmm$ space group.  The $c/a$ ratios have been fixed
to the experimental values in the calculations.  The fractional coordinates for \emph{A} atom
is $(0\,0\,0)$ and for the Fe atom $(0\,\frac{1}{2}\,\frac{1}{4})$.}
\begin{ruledtabular}
\begin{tabular}{c cc c cc c cc}
 & \multicolumn{2}{c}{CaFe$_2$As$_2$} & & \multicolumn{2}{c}{SrFe$_2$As$_2$} &
 & \multicolumn{2}{c}{BaFe$_2$As$_2$} \\
 Cell parameters & Experiment$^{\rm a}$ & GGA    & & Experiment$^{\rm b}$ & GGA & & Experiment$^{\rm c}$ & GGA  \\
 \cline{1-1} \cline{2-3} \cline{5-6} \cline{8-9}
 $a$ (\AA )     &         3.872  & 3.859  & &          3.927 & 3.897   & &      3.964    & 3.924   \\
 $c$ (\AA )     &        11.73   & 11.69  & &         12.37  & 12.27   & &      13.02    & 12.88   \\
 As ($0\,0\,z$) &         0.3612 & 0.3590 & &          0.361 &  0.3526 & &       0.361   &  0.3463 \\
 Fe--As (\AA )  &         2.369  & 2.312  & &          2.396 &  2.320  & &       2.452   &  2.322 \\
 As--Fe--As ($^\circ$)
                &       109.6    & 113.1  & &         110.1  & 114.3   & &     107.8     &  115.4 \\
\end{tabular}
\end{ruledtabular}
\footnotetext[1] {Ref~\onlinecite{wu08}.}
\footnotetext[2] {Ref~\onlinecite{tegel08}.}
\footnotetext[3] {Ref~\onlinecite{rotter08a}.}
\label{tab:gga}
\end{table*}

%
% Conclusions
%
\section{Conclusions}
The detailed temperature dependence of the in-plane infrared active modes has
been studied in \emph{A}Fe$_2$As$_2$ (\emph{A}$\,=\,$Ca, Sr, and Ba) above and below
the structural and magnetic transition at $T_N=172$, 195 and 138~K, respectively.
The phonon frequencies and atomic characters have also been determined from
first principles for each compound in the high-temperature tetragonal phase.
The CaFe$_2$As$_2$ material is the most conventional; the $E_u\rightarrow
B_{2u}+B_{3u}$ splitting into upper and lower branches is observed below $T_N$
for both of the infrared-active modes.  For the low-frequency $E_u$ mode, below
$T_N$ the lower branch continues to soften with decreasing temperature, while
the upper branch hardens.  The high-frequency $E_u$ mode splits into upper and
lower branches of very different strengths, both of which continue to harden
with decreasing temperature in a fashion consistent with an anharmonic decay
scheme.  For both modes, the oscillator strengths are conserved across the
transition, and none of the modes shows any obvious asymmetry.
The behavior of the low-frequency $E_u$ mode is similar in SrFe$_2$As$_2$;
however, while the high-frequency $E_u$ mode is still observed to split in
this material, the upper branch is now extremely weak, and the oscillator
strength of the lower branch increases anomalously below $T_N$.
%
%The SrFe$_2$As$_2$ material shows similar behavior for the low-frequency $E_u$
%splitting below $T_N$; however, for the high-frequency $E_u$ mode only the lower
%branch ($B_{2u}$) is observed \cite{nakajima11}, while the upper branch ($B_{3u}$),
%clearly observed in CaFe$_2$As$_2$, is now extremely weak.
%
This behavior is repeated in BaFe$_2$As$_2$, except that now the oscillator strengths
of the low-frequency modes are no longer conserved; below $T_N$ the high-frequency
$E_u$ mode softens anomalously into a lower branch ($B_{2u}$) \cite{nakajima11} that
increases dramatically in strength with decreasing temperature, while the upper branch
($B_{3u}$) is now entirely absent.
The presence of a SDW (and possibly a weak CDW) along the \emph{b} axis below
$T_N$, suggests that the nature of the bonding along the \emph{a} and \emph{b} axis
has been altered due to orbital ordering, which may lead to an increase in
the strength of the $B_{2u}$ mode and the almost total extinction of the
$B_{3u}$ mode.

%
% acknowledgments
%
%\vspace*{2.0mm}
\begin{acknowledgements}
We would like to acknowledge useful discussions with S.~V.~Dordevic, N.~Felix, and R.~Yang.
Work at the Ames Laboratory (S.~L.~B. and P.~C.~C.) was supported by the U.S. Department of Energy
(DOE), Office of Science, Basic Energy Sciences, Materials Sciences and Engineering Division.
The Ames Laboratory is operated for the U.S. Department of Energy by Iowa State University
under contract No. DE-AC02-07CH11358.  We would like to thanks Alex Thaler and Sheng Ran for
help with sample synthesis.
A.~A. acknowledges funding from The Ambizione Fellowship of the Swiss National Science Foundation.
This work was performed in part at the Aspen Center for Physics, which is supported by National
Science Foundation grant PHY-1066293.
Work at Brookhaven National Laboratory was supported by the Office of Science, U.S. Department
of Energy under Contract No. DE-SC0012704.
\end{acknowledgements}

%
% Appendices
%
\appendix*
\section{Lattice dynamics}
The total energy of \emph{A}Fe$_2$As$_2$ (\emph{A}$\,=\,$Ca, Sr, and Ba) was calculated with the
generalized gradient approximation (GGA) using the full-potential linearized augmented plane-wave
(FP-LAPW) method \cite{singh} with local-orbital extensions \cite{singh91} in the WIEN2k
implementation \cite{wien2k}.  An examination of different Monkhorst-Pack {\em k}-point meshes
indicated that 175 $k$ points ($5\times{5}\times{5}$ mesh) and $R_{mt}k_{max}=8$
was sufficient for good energy convergence.  The geometry of the unit cell was refined
through an iterative process whereby the volume was optimized with respect to the total
energy while the $c/a$ ratio remained fixed.  The atomic fractional coordinates were then
relaxed with respect to the total force, typically resulting in residual forces of less
than 1~mRy/a.u.  This procedure was repeated until no further improvement was obtained.
A comparison of the experimental and calculated (relaxed) unit cell parameters are shown
in Table.~\ref{tab:gga}.  The unit cell parameters are somewhat smaller than the
experimentally-determined values, and the position of the As atom has shifted is closer
to the Fe sheets.

The phonons have been determined using the direct method.  To determine the
phonons at the zone center, a $1\times 1\times 1$ supercell is
sufficient.  To obtain a complete set of Hellmann-Feynman forces, a total of 6
independent displacements are required; because there are always some residual
forces at the atomic sites we have considered symmetric displacements,
which doubles this number, resulting in a total of 12 atomic displacements.
In this case, displacement amplitudes of 0.03~\AA\ were used.  The calculations
have converged when the successive changes for the forces on each atom are
less than 0.1~mRy/au.  The residual forces are collected for each set of
symmetric displacements and a list of the Hellmann-Feynman forces are generated.
Using the program PHONON \cite{phonon} the cumulative force constants deconvoluted
from the Hellmann-Feynman forces are introduced into the dynamical matrix, which
is then diagonalized in order to obtain the phonon frequencies; the atomic intensities
are further calculated to describe the character of the vibration.  The results are
shown in Table~\ref{tab:phonons}.

%
%This work is supported by the Office of Science, U.S. Department of Energy (DOE)
%under Contract No. DE-AC02-98CH10886.

%
%%%%%%%%%%%%%%%%%%%%%%%%%%%%%%%%%%%%%%%%%%%%%%%%%%%%%%%%%%%%%%%%%%%%%%%%%%%%%%
%
% References
%
%\bibliography{references}

\end{document}